\documentclass[english]{IEEEtran}
\pdfoutput=1
\usepackage[T1]{fontenc}
\usepackage[latin9]{inputenc}
\usepackage{amsmath}
\usepackage{graphicx}
\usepackage{setspace}


\usepackage{babel}
\begin{document}

\title{Identifying System-Wide Early Warning Signs of Instability in Stochastic
Power Systems}

\author{\IEEEauthorblockN{Samuel C. Chevalier and Paul D. H. Hines}\thanks{This work was supported by the US DOE, award \#DE-OE0000447, and by the US NSF, award \#ECCS-1254549.} \IEEEauthorblockA{\\College of Engineering and Mathematical Sciences\\ University of Vermont, Burlington, VT}}

\maketitle
\begin{abstract}
Prior research has shown that spectral decomposition of the reduced
power flow Jacobian (RPFJ) can yield participation factors that describe
the extent to which particular buses contribute to particular spectral
components of a power system. Research has also shown that both variance
and autocorrelation of time series voltage data tend to increase as
a power system with stochastically fluctuating loads approaches certain
critical transitions. This paper presents evidence suggesting that
a system's participation factors predict the relative bus voltage
variance values for all nodes in a system. As a result, these participation
factors can be used to filter, weight, and combine real time PMU data
from various locations dispersed throughout a power network in order
to develop coherent measures of global voltage stability. This paper
first describes the method of computing the participation factors.
Next, two potential uses of the participation factors are given: (1)
predicting the relative bus voltage variance magnitudes, and (2) locating
generators at which the autocorrelation of voltage measurements clearly
indicate proximity to critical transitions. The methods are tested
using both analytical and numerical results from a dynamic model of
a 2383-bus test case.\end{abstract}


\section{Introduction}

On a particularly hot day in July of 1987, the power system infrastructure
in Tokyo Japan saw a dramatic increase in demand as millions of air
conditioning units were turned on line. This demand spike occurred
very rapidly and caused system wide voltages to sag. Voltage collapse
soon followed, leaving almost 3 million people without electrical
power. According to \cite{Kurita1988} and \cite{Ohno}, inadequate
operational planning coupled with poor situational awareness were
primary causes of the blackout. Unfortunately, this is not an isolated
voltage collapse incident: in order to optimize limited infrastructure,
many power systems are frequently operated close to critical (or bifurcation)
points, leaving them vulnerable to the devastating effects of voltage
collapse. This paper seeks to improve situational awareness by providing
a method that can use information gleaned from network models to combine
streams of synchrophasor data in ways that provide useful information
about a particular system's proximity to stability limits. Ultimately,
the goal of this work is to provide operators with enhanced tools
for gauging a system\textquoteright s long-term stability. In this
paper, we particularly focus on identifying early warning signs of
\textit{long term voltage stability}, as defined by a joint IEEE PES/CIGRE
task force \cite{Kundur2004}. 

There is increasing evidence that as a dynamical system approaches
a bifurcation, early warning signs (EWSs) of the looming transition
appear in the statistical properties of time series data from that
system. This fact has been evidenced in many complex systems, including
ecological networks, financial markets, the human brain, and power
systems \cite{Ghanavati2014,Scheffer2009}. Researchers have even
found that human depression onset can be predicted by these same statistical
properties \cite{VandeLeemput2014}. In the statistical physics literature
this phenomenon is known as Critical Slowing Down (CSD) \cite{Wissel:1984}.
When stressed, systems experiencing CSD require longer periods to
recover from stochastic perturbations. Two of the most well-documented
signs of CSD are increased variance and autocorrelation \cite{Scheffer2009}. 

Real power systems, particularly as renewable energy production increases,
are constantly subject to stochastically fluctuating supply and demand.
The presence of stochastic power injections has motivated research
to quantify the presence of CSD in bulk power networks, particularly
as early warning signs of bifurcations such as voltage collapse (a
type of Saddle-node bifurcation \cite{Dobson1992}) or oscillatory
instability (a type of Hopf bifurcation). Through simulations, reference
\cite{Cotilla2012} demonstrated that both variance and autocorrelation
in bus voltages increase substantially as several power systems approached
saddle node bifurcation. Similarly, reference \cite{podolsky2013}
computes an auto-correlation function for a power system model to
gauge collapse probability. Finally, variance and autocorrelation
are measured in an unstable power system in \cite{Ghanavati2014}
across many state variables. These results indicate that variance
of bus voltages and autocorrelation of line currents show the most
useful signals of CSD, whereas current angles, voltage angles, generator
rotor angles, and generator speeds did not generally yield sufficiently
strong signs of CSD to provide actionable early warning of bifurcation. 

Although CSD does consistently appear in these systems before bifurcations,
not all variables in a complex system exhibit CSD sufficiently early
to be useful EWSs \cite{Boerlijst2013}. For instance, reference \cite{Ghanavati2014}
destabilized a simulated power system by over stressing all load buses.
Signals were then collected from many nodes in this system, and certain
nodes did not conclusively show measurable early CSD warning signs.
In order to mitigate this problem, we propose a method that combines
measurements from a variety of locations using spectral analysis of
the power flow Jacobian, as introduced in reference \cite{Gao1992}.
By understanding which variables are the best indicators of long-term
voltage stability, we aim to develop measures that are useful for
assessing the stability of an entire system.

It is well known that voltage collapse can cause non-convergence in
AC power flow solvers\cite{Dobson1992}. Additionally, reference \cite{Sauer1990}
shows that only under very strict conditions will the load flow Jacobian
show unambiguous signs of dynamic instability. Since our experiments
do not meet these conditions, spectral analysis of the load flow Jacobian,
as proposed in this paper, does not provide direct warning signs of
dynamic instability. However, since spectral analysis is used in our
approach merely as a means of deciding how to combine PMU data from
diverse locations, our method provides a complement to conventional
approaches that focus only on spectral analysis of the load flow Jacobian.
It is also well known that the saddle-node bifurcation is the maximum
upper limit on system loadability. But this upper limit is seldom
reached, because system dynamics generally become unstable well before
the saddle-node bifurcation occurs \cite{Sauer1990}. Indeed, reference
\cite{Revel2010} shows that a Hopf bifurcation will frequently precede
a saddle-node bifurcation. Therefore, monitoring for voltage instability
is only one important aspect of overall power system stability;there
is a strong link between the occurrence of a Hopf bifurcation and
the proximity of a saddle-node bifurcation.

By combining spectral analysis and CSD theory, this paper shows that
the RPFJ contains valuable information about voltage stability. We
show that participation factors resulting from a spectral analysis
of this matrix can be used to weight and filter real-time PMU data,
thus suggesting a method for combining the data into low-dimensional
metrics of long-term voltage stability. This paper does not seek to
define these metrics in detail; instead, we present a tool (spectral
analysis of the RPFJ) that provides a foundation for the development
of such metrics in future work. Section \ref{sec: Jacobian} of this
paper outlines the mathematical methods and motivation for forming
and decomposing the RPFJ. Section \ref{sec: Experimental_Results}
presents the 2383 test case and shows the potential usefulness of
the participation factors Finally, our conclusions are presented in
Section \ref{sec: Conclusions}.

\section{Spectral Analysis of the Power Flow Jacobian \label{sec: Jacobian}}

This section presents a method for using spectral decomposition of
the RPFJ to identify and weight variables that will most clearly show
evidence of CSD. Further information on this spectral decomposition
approach can be found in \cite{Gao1992}.

The standard power flow Jacobian matrix, which is a linearization
of the steady state power flow equations, is given by (\ref{eq: PF_Jacob}).
\textit{
\begin{equation}
\left[\begin{array}{c}
\Delta\mathbf{P}\\
\Delta\mathbf{Q}
\end{array}\right]=\left[\begin{array}{cc}
J_{\mathbf{P\boldsymbol{\theta}}} & J_{\mathbf{PV}}\\
J_{\mathbf{Q\boldsymbol{\theta}}} & J_{\mathbf{QV}}
\end{array}\right]\left[\begin{array}{c}
\Delta\mathbf{\boldsymbol{\theta}}\\
\Delta\mathbf{V}
\end{array}\right]\label{eq: PF_Jacob}
\end{equation}
}

In order to perform V-Q sensitivity analysis (an important aspect
of voltage stability analysis), we assume that the incremental change
in real power $\Delta\mathbf{P}$ is equal to 0. In this way, we can
study how incremental changes in injected reactive power affect system
voltages. Setting $\Delta\mathbf{P}=\mathbf{0}$ and rearranging terms
to remove $\Delta\boldsymbol{\theta}$, the expression for the reduced
Jacobian is defined:
\begin{equation}
\Delta\mathbf{Q}=\left[J_{\mathbf{QV}}-J_{\mathbf{Q}\boldsymbol{\theta}}J_{\mathbf{P}\boldsymbol{\theta}}^{-1}J_{\mathbf{PV}}\right]\Delta\mathbf{V}=\left[J_{R}\right]\Delta\mathbf{V}\label{eq: dQ_dV}
\end{equation}

Assuming that Newton-Raphson converges to a power flow solution for
the system being studied, the matrix $J_{R}$ is non-singular and
can be written as the product of its right eigenvector matrix $R$,
its left eigenvector matrix $L$, and its diagonal eigenvalue matrix
$\Lambda$, such that:
\begin{equation}
J_{R}=R\Lambda L\label{eq: Jr_Diagonal}
\end{equation}

The left and right eigenvectors can then be orthonormalized such that,
for the right eigenvector $\mathbf{r}_{i}$ (column vector) and the
left eigenvector $\mathbf{l}_{j}$ (row vector), the Konecker delta
function defines their relationship:
\begin{equation}
\mathbf{l}_{j}\mathbf{r}_{i}=\delta_{j,i}=\begin{cases}
1 & i=j\\
0 & i\neq j
\end{cases}
\end{equation}

\noindent We begin our method by decomposing $J_{R}$ using a simple
similarity transform. The transform is substituted into (\ref{eq: dQ_dV}):\textit{\small{}
\begin{eqnarray}
 &  & \Delta\mathbf{Q}=\nonumber \\
 &  & \left[\begin{array}{c}
\mathbf{r}_{1}\end{array}\begin{array}{ccc}
\mathbf{r}_{2} & \cdots & \mathbf{r}_{n}\end{array}\right]\left[\begin{array}{cccc}
\lambda_{1} & 0 & \cdots & 0\\
0 & \lambda_{2} &  & \vdots\\
\vdots &  & \ddots & 0\\
0 & \cdots & 0 & \lambda_{n}
\end{array}\right]\left[\begin{array}{c}
\mathbf{l}_{1}\\
\mathbf{l}_{2}\\
\vdots\\
\mathbf{l}_{n}
\end{array}\right]\Delta\mathbf{V}\nonumber \\
 & = & \left[\begin{array}{cccc}
r_{1,1} & r_{2,1} & \cdots & r_{n,1}\\
r_{1,2} & r_{2,2} &  & r_{n,2}\\
\vdots &  & \ddots & \vdots\\
r_{1,n} & r_{2,n} & \cdots & r_{n,n}
\end{array}\right]\left[\begin{array}{c}
\lambda_{1}(\mathbf{l}_{1}\cdot\Delta\mathbf{V})\\
\lambda_{2}(\mathbf{l}_{2}\cdot\Delta\mathbf{V})\\
\vdots\\
\lambda_{n}(\mathbf{l}_{n}\cdot\Delta\mathbf{V})
\end{array}\right]\label{eq: del_Q_Mat}
\end{eqnarray}
}The effect of this transform can be made more clear by investigating
how changing voltage affects the change in injected reactive power
of a single bus ($\triangle Q_{1}$ for example). This is shown in
(\ref{eq: del_Q1_1}).
\begin{eqnarray}
\triangle Q_{1} & = & r_{1,1}\lambda_{1}(\mathbf{l}_{1}\cdot\Delta\mathbf{V})+r_{2,1}\lambda_{2}(\mathbf{l}_{2}\cdot\Delta\mathbf{V})+\label{eq: del_Q1_1}\\
 &  & \cdots+r_{n,1}\lambda_{n}(\mathbf{l}_{n}\cdot\Delta\mathbf{V})\nonumber 
\end{eqnarray}

\noindent In order to determine how the reactive power at bus $i$
is affected by the voltage at only bus $i$, we simply hold all other
voltage magnitudes constant. If we choose $i=1$, the voltage differential
vector becomes $\Delta\mathbf{V}=[\begin{array}{cccc}
\triangle V_{1} & 0 & \cdots & 0\end{array}]$. The reactive power differential equation changes accordingly.
\begin{equation}
\triangle Q_{1}=\left(\lambda_{1}r_{1,1}l_{1,1}+\lambda_{2}r_{2,1}l_{2,1}+\cdots+\lambda_{n}r_{n,1}l_{n,1}\right)\triangle V_{1}\label{eq:  del_Q1_2}
\end{equation}

\noindent At this point, we can define and incorporate the participation
factors. The indices in the following equation refer to the $j^{th}$
row and the $i^{th}$ column of the right eigenvector matrix $R$
and the $i^{th}$ row and the $j^{th}$ column of the left eigenvector
matrix $L$.
\begin{equation}
\rho_{i,j}=R_{j,i}L_{i,j}\label{eq: PF_Eq}
\end{equation}

\noindent Therefore, $\rho_{i,j}$ defines how the $j^{th}$ state
is affected by the $i^{th}$ eigenvalue. Clearly, individual reactive
power states can be expressed as a superposition of eigenvalues of
varying degrees of participation. If we compute the reactive power
changes at each bus based on the voltage changes at each corresponding
local bus, we obtain the following set of equations.
\[
\triangle Q_{1}=\left(\lambda_{1}\rho_{1,1}+\lambda_{2}\rho_{2,1}+\cdots+\lambda_{n}\rho_{n,1}\right)\triangle V_{1}
\]
\[
\triangle Q_{2}=\left(\lambda_{1}\rho_{1,2}+\lambda_{2}\rho_{2,2}+\cdots+\lambda_{n}\rho_{n,2}\right)\triangle V_{2}
\]
\[
\vdots
\]
\[
\triangle Q_{n}=\left(\lambda_{1}\rho_{1,n}+\lambda_{2}\rho_{2,n}+\cdots+\lambda_{n}\rho_{n,n}\right)\triangle V_{n}
\]

In these equations, a reactive power state is expressed as a superposition
of eigenvalues. Conversely, we can also express each eigenvalue as
a superposition of different state contributions. The reason why such
an expression is useful is shown through (\ref{eq: Lambda_sum}).
Recognizing that $R=L^{-1}$, the following manipulations may be made.\textit{\small{}
\begin{eqnarray*}
 &  & \Delta\mathbf{Q}=\\
 &  & \left[\begin{array}{c}
\mathbf{r}_{1}\end{array}\begin{array}{ccc}
\mathbf{r}_{2} & \cdots & \mathbf{r}_{n}\end{array}\right]\left[\begin{array}{cccc}
\lambda_{1} & 0 & \cdots & 0\\
0 & \lambda_{2} &  & \vdots\\
\vdots &  & \ddots & 0\\
0 & \cdots & 0 & \lambda_{n}
\end{array}\right]\left[\begin{array}{c}
\mathbf{l}_{1}\\
\mathbf{l}_{2}\\
\vdots\\
\mathbf{l}_{n}
\end{array}\right]\Delta\mathbf{V}
\end{eqnarray*}
}{\small \par}

\textit{\small{}
\[
\left[\begin{array}{c}
\mathbf{l}_{1}\\
\mathbf{l}_{2}\\
\vdots\\
\mathbf{l}_{n}
\end{array}\right]\Delta\mathbf{Q}=\left[\begin{array}{cccc}
\lambda_{1} & 0 & \cdots & 0\\
0 & \lambda_{2} &  & \vdots\\
\vdots &  & \ddots & 0\\
0 & \cdots & 0 & \lambda_{n}
\end{array}\right]\left[\begin{array}{c}
\mathbf{l}_{1}\\
\mathbf{l}_{2}\\
\vdots\\
\mathbf{l}_{n}
\end{array}\right]\Delta\mathbf{V}
\]
}{\small \par}

\noindent Now, we can isolate a single eigenvalue ($\lambda_{1}$,
for example).
\begin{equation}
\mathbf{l}_{1}\Delta\mathbf{Q}=\lambda_{1}\mathbf{l}_{1}\Delta\mathbf{V}
\end{equation}
\begin{eqnarray}
 &  & l_{1,1}\triangle Q_{1}+l_{1,2}\triangle Q_{2}+\cdots+l_{1,n}\triangle Q_{n}=\\
 &  & \lambda_{1}\left(l_{1,1}\triangle\mathrm{V}_{1}+l_{1,2}\triangle\mathrm{V}_{2}+\cdots+l_{1,n}\triangle\mathrm{V}_{n}\right)\nonumber 
\end{eqnarray}

\noindent Clearly, the relationship between $\triangle Q$ and $\triangle\mathrm{V}$
for the $j^{th}$ isolated state (holding all else constant) is given
by the following expression.
\begin{equation}
\frac{q_{1,j}\triangle\mathrm{V}_{j}}{q_{1,j}\triangle Q_{j}}=\frac{\triangle\mathrm{V}_{j}}{\triangle Q_{j}}=\frac{1}{\lambda_{j}}\label{eq: dV_dQ_Lambda}
\end{equation}

\noindent This is true for all states of a given eigenvalue. Therefore,
the spectral component that will have the largest voltage variation
for a given reactive power change will have the smallest eigenvalue.
For this reason, the participation factors of this eigenvalue will
be of great interest to study. The $j^{th}$ eigenvalue can be written
as a summation of $n$ unique states. In this way, (\ref{eq: Lambda_sum})
shows how each state participates in the $j^{th}$ eigenvalue of a
system.
\begin{equation}
\lambda_{j}=\lambda_{j}\rho_{j,1}+\lambda_{j}\rho_{j,2}+\cdots+\lambda_{j}\rho_{j,n}\label{eq: Lambda_sum}
\end{equation}

There are many different ways to use the eigenvalues and eigenvectors
of $J_{R}$. For instance, \cite{Gao1992} suggest using the smallest
eigenvalue of $J_{R}$ to gauge proximity to bifurcation. Such stability
analysis, though, is based solely on the decomposition of a model
based static matrix and is highly limited in nature, as outlined by
Pal in the discussion section of \cite{Gao1992}. Instead, we propose
that $J_{R}$ can be leveraged as tool to combine and thus interpret
streams of PMU data. Detecting Critical Slowing Down in time series
data is a purely data driven stability assessment, but it can be difficult
to understand which nodes will show the strongest EWSs \cite{Ghanavati2014}.
Therefore, the novel approach outlined in this paper uses results
from the static decomposition above to weight and interpret incoming
dynamic data.

\section{Experimental results: Using participation factors to combine PMU
data\label{sec: Experimental_Results}}

\subsection{Polish Test Case Overview \label{sub: System Overview}}

In order to test our methods, we use data from the 2383-bus dynamic
Polish test system. This network contains 327 four-variable synchronous
generators. Each generator is equipped with a three-variable turbine
governor model for frequency control and a four-variable exciter model
(AVR) for voltage regulation. There are 322 shunt loads (all connected
to generator buses) and 1503 active and reactive loads spread throughout
the system. In order to push the system towards voltage collapse,
we employed a simple uniform loading of all loads (except for those
attached to generator buses). This method is justified in \cite{Kundur2004}.
Half of the PQ bus loads are modeled as voltage controlled loads,
while the other half are modeled as frequency controlled loads. Parameters
controlling the voltage controlled loads are modeled after the Nordic
Test System in \cite{AndradedosSantos2015}, while parameters controlling
the frequency controlled loads are modeled after the 39 bus test system
described in \cite{Ghanavati2014}.

As in \cite{Ghanavati2014}, we model this larger network using a
set of stochastically forced differential algebraic equations, which
can be written as:
\begin{equation}
\dot{\mathbf{x}}=\mathbf{f}(\mathbf{x},\mathbf{y})\label{eq:Diff_Eqs}
\end{equation}
\begin{equation}
\mathbf{0}=\mathbf{g}(\mathbf{x},\mathbf{y},\mathbf{u})\label{eq:Alg_Eqs}
\end{equation}
where $\mathbf{f}$ and $\mathbf{g}$ represent the differential and
algebraic equations governing the system, $\mathbf{x}$ and $\mathbf{y}$
are the differential and algebraic variables of these equations, and
\textbf{$\mathbf{u}$} represents stochastic power (load or supply)
fluctuations. $\mathbf{u}$ follows a mean-reverting Ornstein-Uhlenbeck
process:
\begin{equation}
\mathbf{\dot{u}}=-E\mathbf{u}+\mathbf{\boldsymbol{\xi}}\label{eq:O-U_Process}
\end{equation}
where $E$ is a diagonal matrix whose diagonal entries equal the inverse
correlation times $t_{corr}^{-1}$ of load fluctuations and $\mathbf{\boldsymbol{\xi}}$
is a vector of zero-mean independent Gaussian random variables. A
further description of our noise model can be found in Sec. II A of
\cite{Ghanavati2014}. Also given in \cite{Ghanavati2014} is a method
for analytically computing the covariance and correlation matrices
for all state and algebraic variables. We used this method to pre-compute
the variance of voltages in the 2383-bus Polish system. After thorough
testing, we found the analytically-calculated covariance and correlation
matrices to be just as accurate on the large Polish system as they
were on the small 39 bus system. Thus, the data presented in the following
two sections use the analytically calculated results rather than averaged
dynamic simulation results.

In order to push the system towards a critical transition (voltage
collapse), we increase all loads and generator set points by a constant
loading factor $b$, which ranges from $b=1$ up to $b=1.92$. We
empirically found that voltage collapse occurs when the load factor
increases past $b=1.923$.

The concept of a limit-induced bifurcation is an important topic discussed
in \cite{Lerm1998}. Power system limits, such as reactive power generation
limits, are an important aspect of stability analysis. However, in
order to focus our analysis on voltage collapse without the possibility
of additional bifurcations due to limits, we increased limits in our
test case so that the system can run up to $b=1.92$ without hitting
a limit-induced bifurcation. This simplification allows us to focus
our study on the effects of pure voltage collapse. Future work will
extend our method to the case of other types of bifurcations.

\subsection{Evidence for Bus Voltage Variance Prediction \label{sub: Var_Predictions}}

\begin{figure}
\noindent \begin{centering}
\includegraphics[width=1\textwidth,height=0.28\textheight,keepaspectratio]{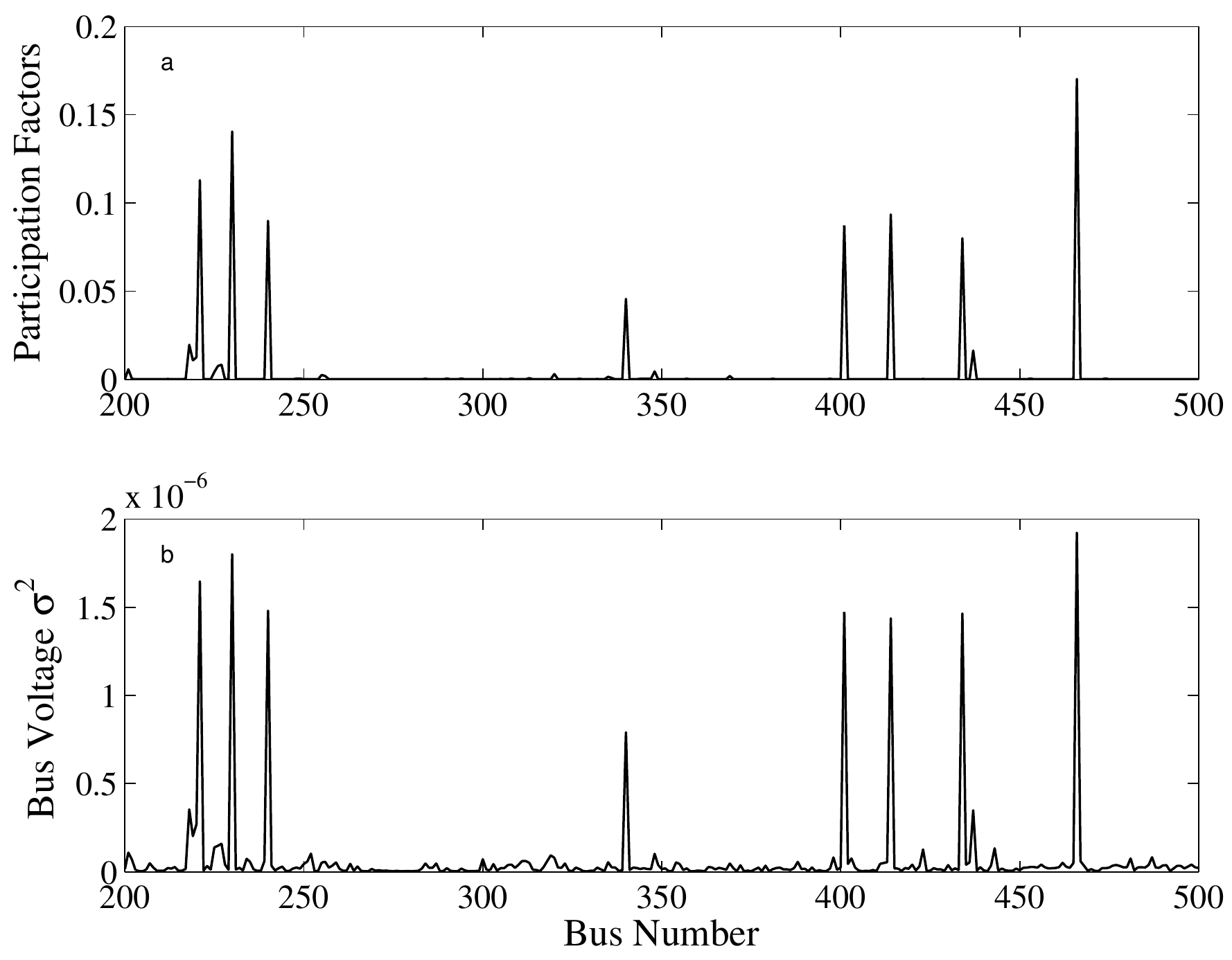}
\par\end{centering}

\protect\caption{\label{fig:Var_vs_PF}Participation factors and voltage variance values
for buses 200 through 500 from the loaded 2383 bus system. Bus voltage
variances (below) correlate almost perfectly with the participation
factors of the smallest eigenvalue of the RPFJ (above). }
\end{figure}

As indicated by (\ref{eq: dV_dQ_Lambda}), the smallest eigenvalue
of $J_{R}$ corresponds to the spectral component that will yield
the largest voltage variation for a given variation in reactive power.
When the participation factors corresponding to the smallest eigenvalue
are plotted, they are shown to directly predict the relative bus voltage
variance strengths. Fig. \ref{fig:Var_vs_PF} shows two plots. The
top plot corresponds to the participation factors for buses 200 through
500, and the bottom plot shows the true bus voltage variance, derived
analytically, for buses 200 through 500. The remaining system buses
are left out for the sake of clarity. Despite the fact that the participation
factors are completely blind to the dynamics of the system, they are
still quite successful at predicting the relative variance strengths.

\noindent 

As shown in reference \cite{ghanavati2013understanding}, increasing
voltage variance is due to buses that are operating closer to the
limit along the PV curve. Therefore, participation factors of the
smallest eigenvalue also identify the node voltages which, as the
system is overloaded, begin to diverge from their nominal values.
These tend to be the nodes that are primarily responsible for non-convergence
in the power flow equations. Interestingly, as PQ buses in the system
are increasingly loaded, the recalculated participation factors do
not change drastically (for a uniform loading condition). This is
equivalent to saying that the \emph{spectral components} do not change
significantly. This is a useful result, since state-estimator derived
power flow models are only typically computed periodically during
power systems operations. 

\begin{figure}
\noindent \begin{centering}
\includegraphics[width=1\columnwidth,height=0.27\textheight,keepaspectratio]{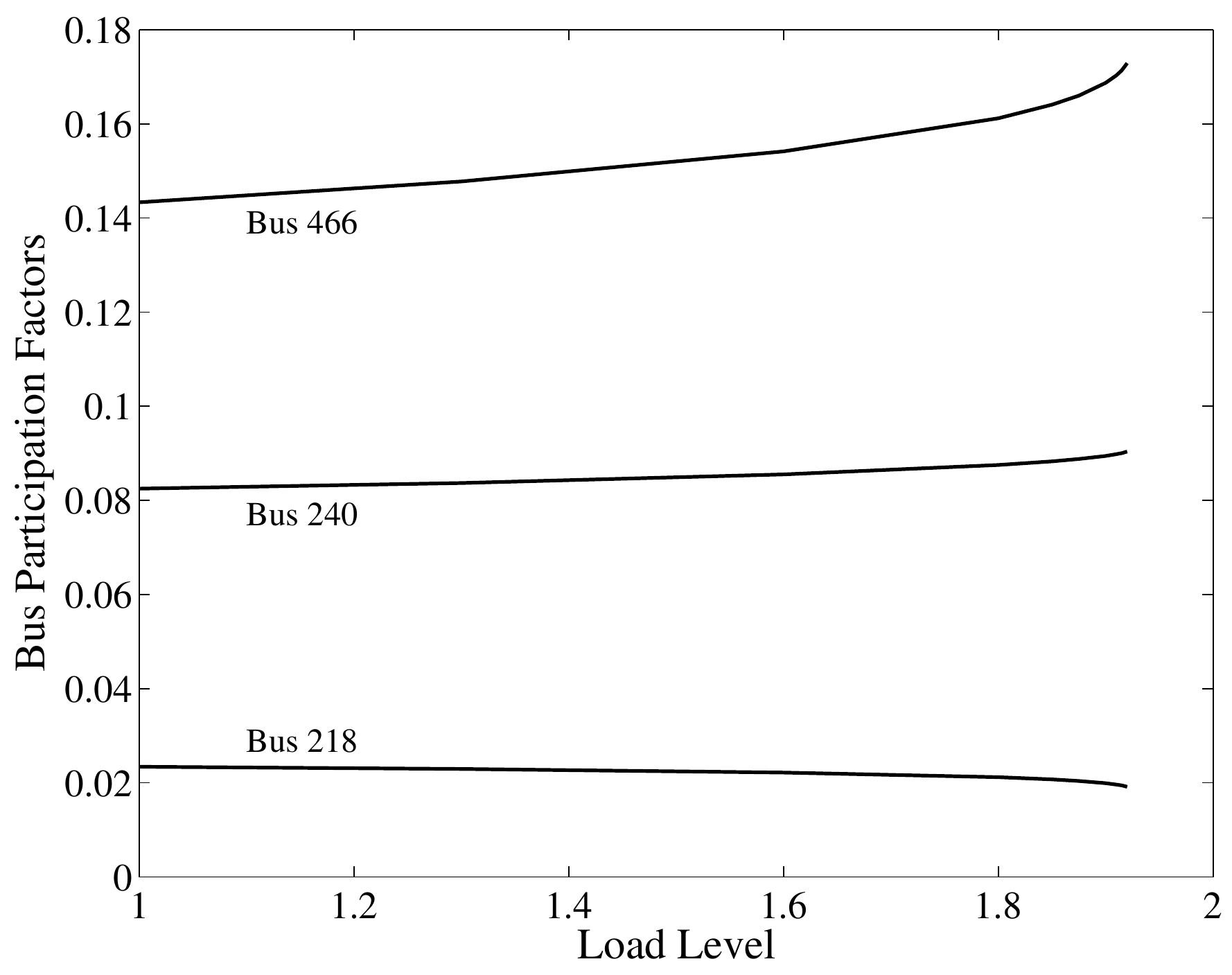}
\par\end{centering}

\protect\caption{\label{fig:PF_Increases}Participation factors for three different
buses at different load levels. As the system is increasingly loaded
(right up to bifurcation), bus 466 (the most unstable bus) sees a
sharp increase in participation to the instability. Bus 240 (the 5th
most unstable bus) sees a very slight increase, while bus 218 (the
10th most unstable bus) begins to see a decrease.}
\end{figure}

As indicted previously, participation factors of the most unstable
nodes provide a very clear indication of the relative bus voltage
variance strengths. Therefore, as the system is increasingly loaded,
the most unstable nodes will begin to have larger and larger participation
factors as their relative variance strengths grow relative to other,
more stable nodes. Fig.~\ref{fig:PF_Increases} shows an example
of this for the 2383 bus system. As the system is loaded, the relative
strength of the most unstable bus' participation increases almost
linearly, but when the critical transition approaches, the participation
begins to climb more steeply.

\subsection{Evidence for Locating Generators with Elevated Voltage Autocorrelation
\label{sub: AutoCorr_Gens}}

CSD theory predicts that signals from a system approaching a critical
transition will show increasing levels of auto-correlation, $R(\Delta t)$.
This can be due to the system's reduced ability to respond to high
frequency fluctuations \cite{Dakos2012}, but the system also begins
to return to an equilibrium state more slowly after perturbations
\cite{ghanavati2013understanding}. In a power system, system-wide
increases inautocorrelation are typically indicators of increasingly
unstable generator dynamics. These dynamics are driven by the load
variations, since this is where the noise is being injected.

Fig. \ref{fig:Var_vs_PF}, clearly shows that a small number of buses
in our test case have particularly high voltage variances measurements.
Looking at the topology of this network, we find that these nodes
are in fact separated by only a small number of transmission lines
suggesting that these buses represent a weak load pocket. The participation
factors are thus useful for identifying load pockets. Many of the
buses connected to this pocket show high variance, and are therefore
driving the autocorrelation of the most proximal generators. By identifying
generator proximity to unstable loads, the autocorrelation of the
output signals (voltage and current) of close generators can be scrutinized.

\begin{figure}
\noindent \begin{centering}
\includegraphics[scale=0.6]{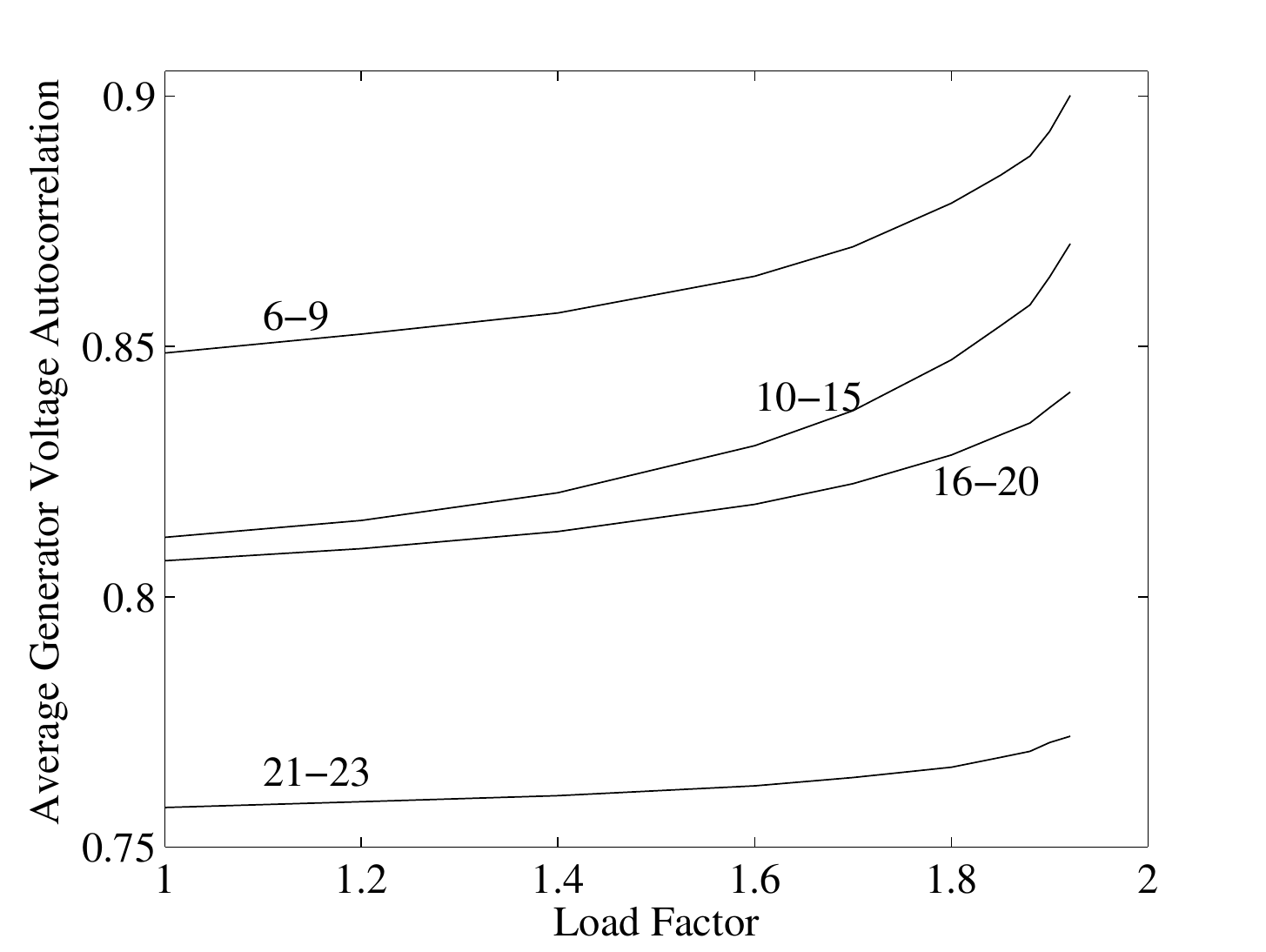}
\par\end{centering}

\protect\caption{\label{fig: xcorr_increases} Average autocorrelation of voltage magnitude
measurements for several different groups of generators at varying
load levels. The top trace is all generators located between 6 and
9 transmission lines (hops) from the load pocket. The second groupincludes
generators 10 and 15 transmission lines distant from the load pocket.
The third group includes generators 16 and 20 lines from the pocket,
and the fourth group is generators between 21 and 23 lines distant.}
\end{figure}

To study this further, the 327 generators of the Polish system were
group according to their distance (quantified by line count) from
the load pocket center. Next, the system wide algebraic correlation
matrix was derived for a series of increasing load parameters. For
each generator grouping, the average bus voltage autocorrelation value
was computed and plotted. Fig. \ref{fig: xcorr_increases} shows the
average voltage autocorrelation, with a time lag of $\triangle t=.2$s,
for several different group of generators and varying load levels.
(See \cite{ghanavati2013understanding} for a discussion of this choice
for $\Delta t$.).The generators are grouped (and identified by) their
proximity to the load pocket shown in Fig. \ref{fig:Var_vs_PF}. Clearly,
the generators that are closest to this load pocket show the largest
average autocorrelation statistics. This is a very useful result.

\noindent 

\section{Conclusions\label{sec: Conclusions}}

This paper presents evidence that participation factors from a spectral
decomposition of the reduced power flow Jacobian can be used to design
methods for combining sychrophasor measurements to produce system-wide
indicators of instability in power systems. Our approach uses spectral
information from the power flow Jacobian, which can be updated every
few minutes through the SCADA network in combination with high sample-rate
voltage magnitude measurements, which can be collected from synchronized
phasor measurement systems deployed throughout the system. The results
suggest that that a combination of a power flow model and streaming
PMU data analysis can be used to be used to develop system wide stability
metrics. The detailed development of these metrics is a topic for
future research.

\bibliographystyle{IEEEtran}
\bibliography{PES_Paper}

\section*{Author biographies}

\begin{singlespace}
\noindent \textbf{\small{}Paul D. H. Hines}{\small{} (SM`14) received
the Ph.D. in Engineering and Public Policy from Carnegie Mellon U.
in 2007 and M.S. (2001) and B.S. (1997) degrees in Electrical Engineering
from the U. of Washington and Seattle Pacific U., respectively. He
is currently Associate Professor in Electrical Engineering at the
U. of Vermont.}\\
{\small \par}

\noindent \textbf{\small{}Samuel C. Chevalier}{\small{} (S`13) received
a B.S. in Electrical Engineering from the University of Vermont in
2015. He is currently pursuing an M.S. degree in Electrical Engineering
from UVM, and his research interests include stochastic power system
stability and Smart Grid.}\end{singlespace}

\end{document}